\documentclass[pre,showpacs,floatfix,twocolumn,10pt]{revtex4-1}

\usepackage{amssymb,amsfonts,amsmath}
\usepackage{bm}				% bold math symbols
\usepackage{latexsym}
\usepackage{graphicx}

\usepackage[usenames]{color}
\usepackage{soul}

\newcommand{\be}{\begin{equation}}
\newcommand{\ee}{\end{equation}}

\newcommand{\ben}{\begin{eqnarray}}
\newcommand{\een}{\end{eqnarray}}

\renewcommand{\vec}[1]{{\bf {#1}}}

\newcommand{\vrr}{{\bf{r}}}

\def\(({\left(}
\def\)){\right)}
\def\[[{\left[}
\def\]]{\right]}

\begin{document}

\title{Scaling laws and bulk-boundary decoupling in heat flow}

\author{Jes\'us J. del Pozo}
\email{jpozo@onsager.ugr.es}
\affiliation{Institute Carlos I for Theoretical and Computational Physics and Departamento de Electromagnetismo y F\'isica de la Materia, Universidad de Granada, 18071 Granada, Spain}
\author{Pedro L. Garrido}
\email{garrido@onsager.ugr.es}
\affiliation{Institute Carlos I for Theoretical and Computational Physics and Departamento de Electromagnetismo y F\'isica de la Materia, Universidad de Granada, 18071 Granada, Spain}
\author{Pablo I. Hurtado}
\email{phurtado@onsager.ugr.es}
\affiliation{Institute Carlos I for Theoretical and Computational Physics and Departamento de Electromagnetismo y F\'isica de la Materia, Universidad de Granada, 18071 Granada, Spain}

\date{\today}


\begin{abstract} 
When driven out of equilibrium by a temperature gradient, fluids respond by developing a nontrivial, inhomogeneous structure according to the governing macroscopic laws. Here we show that such structure obeys strikingly simple scaling laws arbitrarily far from equilibrium, provided that both macroscopic local equilibrium and Fourier's law hold. Extensive simulations of hard disk fluids confirm the scaling laws even under strong temperature gradients, implying that Fourier's law remains valid in this highly nonlinear regime, with putative corrections absorbed into a nonlinear conductivity functional. In addition, our results show that the scaling laws are robust in the presence of strong finite-size effects, hinting at a subtle bulk-boundary decoupling mechanism which enforces the macroscopic laws on the bulk of the finite-sized fluid. This allows to measure for the first time the marginal anomaly of the heat conductivity predicted for hard disks.
\end{abstract}

\maketitle

The understanding of nonequilibrium behavior remains as one of the major challenges in theoretical physics, even in the simplest situations posed by nonequilibrium steady states (NESSs) \cite{noneq,Bertini,Derrida,Touchette,Pablo,IFR,GC,ECM}. The first thing one notices in typical NESSs (as those obtained for fluids under a temperature gradient) is the nontrivial, inhomogeneous structure that the system of interest develops in response to the nonequilibrium driving. This structure, readily measurable in experiments or simulations, carries information on the governing nonequilibrium macroscopic laws (e.g. Fourier's law) which emerge from the myriad of interacting microscopic constituents. It is therefore of paramount importance to understand general properties of these structures, consubstantial to nonequilibrium behavior. With this idea in mind, we derive here a set of simple yet general scaling laws for a broad class of $d$-dimensional fluids driven far from equilibrium by a temperature gradient. In particular, we show that the fluid's density and temperature profiles follow from two master curves, independent of the driving force and the system parameters, after a simple linear scaling of space. This strong result is based on two mild hypotheses, namely macroscopic local equilibrium and Fourier's law, together with a rather general assumption on the fluid's equation of state. 

\begin{figure}[b]
\vspace{-0.3cm}
\centerline{\includegraphics[width=9.cm]{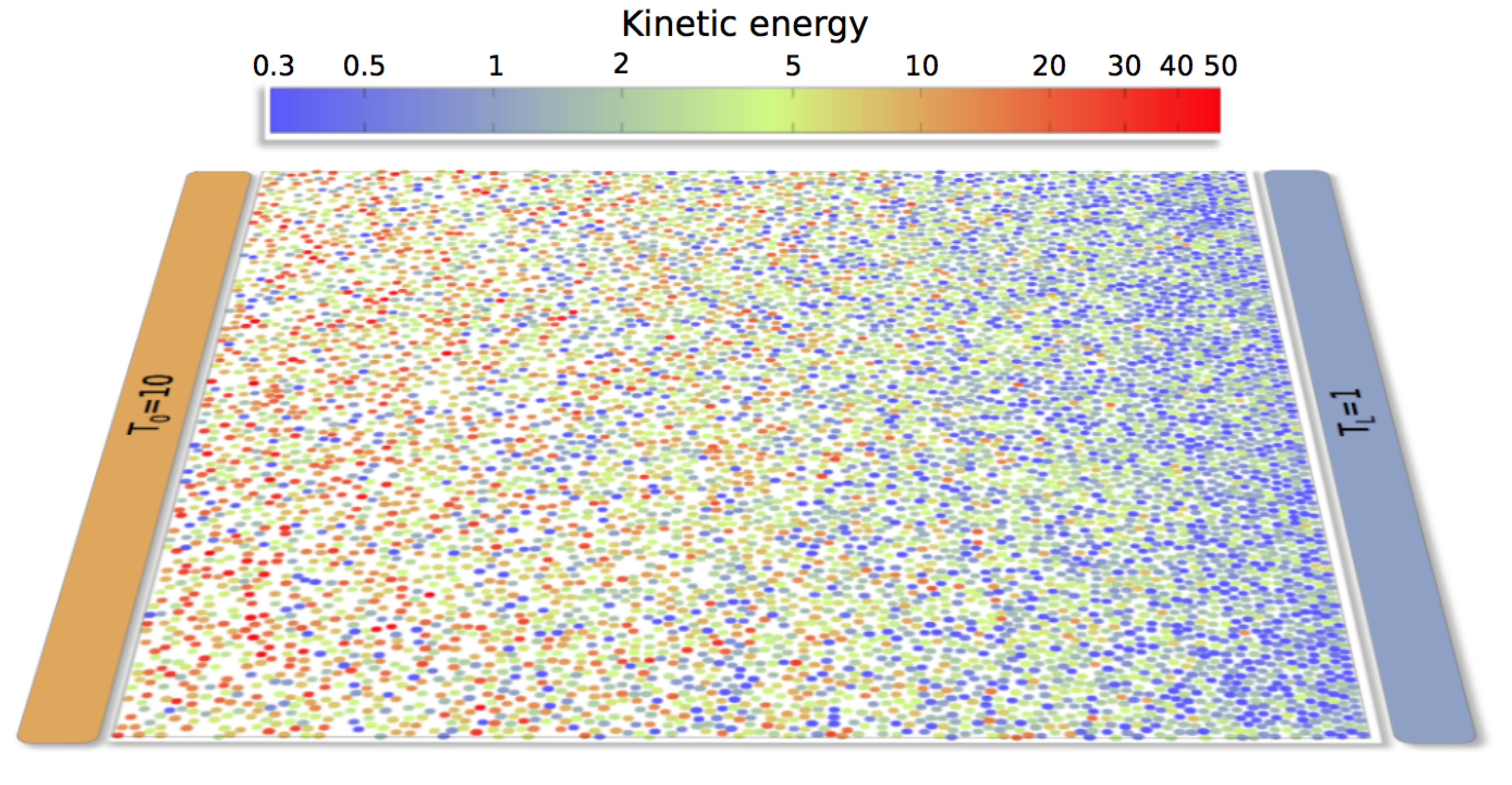}}
\caption{\small (Color online)
Snapshot of a typical configuration with $N=7838$ hard disks at $\eta=0.5$, subject to a temperature gradient ($T_0=10$, $T_L=1$).
 Colors represent kinetic energy.
}
\label{fig1}
\end{figure}

We then proceed to test the emerging picture in a quintessential model, the hard disk fluid. Hard sphere (HS) models and their relatives are among the most successful, inspiring and prolific models of physics, as they contain the essential ingredients to understand a large class of complex phenomena, from phase transitions or heat transport to glassy dynamics, jamming, or the physics of liquid crystals and granular materials, to mention just a few \cite{Mulero,Chaikin,soft,glass,deGennes,granular,tails,Alder,Rosenbluth,IFR,fourier1,fourier2,shearCL,Pedro,chaos}, turning general results for these systems even more appealing. Extensive computer simulations of hard disks under temperature gradients confirm the above scaling laws with surprising accuracy, showing that Fourier's law remains valid for each $N$ even under strong gradients and despite the marginally divergent heat conductivity of hard disks (which has however minor numerical consequences \cite{tails,fourier1}). This proves that, at least for hard disks under quiescent heat transport, the putative higher-order corrections to Fourier's law can be accounted for by a nonlinear conductivity functional, see below. Our results also reveal a striking decoupling between the bulk fluid, which behaves \emph{macroscopically}, and two boundary layers near the thermal walls, which sum up all sorts of artificial finite-size and boundary corrections to renormalize the effective  boundary conditions on the remaining bulk. This bulk-boundary decoupling phenomenon, which probably characterizes the physics of a large class of fluids, allows to obtain reliable measurements of collective properties of macroscopic systems using data from finite-size simulations. We illustrate this idea by measuring the hard-disks heat conductivity for a broad range of densities, confirming for the first time its marginally anomalous $\sqrt{\ln N}$-behavior in the large size limit as a result of the long-time tails \cite{tails}. This shows that our scaling method keeps physically relevant finite-size information while getting rid of artificial finite-size and boundary corrections. 

We hence consider a $d$-dimensional fluid in a box of linear size $L$ and global packing fraction $\eta=Nv/L^d$, with $v$ the volume of a fluid's particle, driven out of equilibrium by two boundary heat baths (say along the $x$-direction) operating at different temperatures, $T_0>T_L$, see e.g. Fig. 1. Our results below are based on two simple hypotheses, namely (i) Local Equilibrium (LE) and (ii) Fourier's law. In particular, with {(i)} we assume that LE holds at the macroscopic level, in the sense that the \emph{local} density and temperature are related by the \emph{equilibrium} equation of state (EoS) $Q=q(\rho,T)$, with $Q=Pv$ and $P$ the pressure. This hypothesis has been recently shown to hold empirically for hard disks under a broad range of temperature gradients \cite{jpozo1}. On the other hand, Fourier's law states that, in the steady state, the heat current $J$ is proportional to the temperature gradient \cite{fourier1,fourier2}, i.e.
\be
J=-\kappa(\rho,T)\frac{dT(x)}{dx}\, , \quad x\in[0,L] \, ,
\label{fourier}
\ee
where $\kappa(\rho,T)$ is the thermal conductivity, that may depend in general on the local temperature $T(\vrr)$ and on the local packing fraction $\rho(\vrr)$. Fourier's law (\ref{fourier}) formally applies in the limit of small temperature gradients, with higher-order (Burnett) corrections in the gradient conjectured for stronger driving \cite{Mulero}. However, our results below suggest that, at least for quiescent heat transfer, these corrections are absorbed into a nonlinear conductivity functional, extending the validity of Fourier's law deep into the strongly nonlinear regime. 

Interestingly, we may use now macroscopic LE to write Fourier's law in terms only of the density field. To do so, we need the EoS to be invertible in the $(\rho,T)$-range of interest, an assumption which holds valid for most fluids away from a critical point. In this case, inverting the EoS $Q=q(\rho,T)$ yields $T=f_Q(\rho)$, with $f_Q(\rho)$ an uniparametric curve such that $q[\rho,f_Q(\rho)]=Q$. Similarly, the heat conductivity follows as $\kappa(\rho,T)=\kappa[\rho,f_Q(\rho)]\equiv k_Q(\rho)$, defining another uniparametric function $k_Q(\rho)$. This allows to rewrite Fourier's law  (\ref{fourier}) as
\be
J=G'_Q(\rho)\frac{d\rho}{dx} = \frac{dG_Q(\rho)}{dx} \, ,
\label{JQcte}
\ee
where $G'_Q(\rho)\equiv -k_Q(\rho) f'_Q(\rho)$ and $'$ denotes derivative with respect to the argument. This equation, together with the boundary conditions for the density field \cite{conds}, completely define the macroscopic problem in terms of $\rho(\vrr)$. A striking consequence of hypotheses (i)-(ii) can be now directly inferred from eq. (\ref{JQcte}). In fact, as both $J$ and $Q$ are state-dependent constants, this immediately implies that $G_Q[\rho(x)]=Jx +\zeta$, i.e. $G_Q[\rho(x)]$ is a linear function of position, with slope $J$ and $\zeta=G_Q(\rho_0)$ an arbitrary constant, or equivalently \cite{exist}
\be
\rho(x)=G_Q^{-1}(Jx+\zeta) \, .
\label{density}
\ee
Therefore, there exists a single master \emph{surface} $\bar{\rho}_Q(y)\equiv G_Q^{-1}(y)$ in $y-Q$ space from which any steady state density profile follows after a linear spatial scaling $x=(y-\zeta)/J$. Furthermore, this scaling behavior is transferred to temperature profiles via the local EoS, which yields another master surface $\bar{T}_Q(y)=f_Q[G_Q^{-1}(y)]$. These scaling laws, that completely characterize heat flow in the system of interest, are independent of the packing fraction $\eta$ or the nonequilibrium driving defined by the baths temperatures $T_0$ and $T_L$, depending exclusively on the uniparametric functions $f_Q(\rho)$ and $k_Q(\rho)$ controlling the system macroscopic behavior. Alternatively, eq.~(\ref{density}) implies that any measured steady density profile can be collapsed onto the master surface $\bar{\rho}_Q(y)$ by scaling space by the associated current $J$ and shifting the resulting profile an arbitrary constant $\zeta$ (similarly for temperature profiles). This suggests a simple scaling method to obtain the master curves in simulations and experiments that we exploit below.

For systems with homogeneous interparticle potentials, $V(\vrr)\propto r^{-n}$, both the EoS and the heat conductivity exhibit a well-known density-temperature separability (see Appendix A) \cite{Dyre}, which simplifies the form of the general scaling laws derived above. In particular, for hard disks the EoS takes the simpler form $Q=T\, q(\rho)$, with $q(\rho)$ an unknown function for which many accurate approximations can be found in literature \cite{Mulero,Santos,jpozo1}. The conductivity also takes the separable form $\kappa(\rho,T)=\sqrt{T}\, k(\rho)$, where again $k(\rho)$ is still unknown. A reasonably good approximation is obtained however from Enskog kinetic theory for hard disks \cite{enskog,Gass,brey}. It is then easy to show that, in this case, the above master surfaces collapse onto a pair of universal \emph{curves}. In particular, for hard disks $G_Q(\rho)=Q^{3/2} G(\rho)$, with $G'(\rho)\equiv k(\rho) q(\rho)^{-5/2} q'(\rho)$, so all density profiles scale as $\rho(x)=G^{-1}(\psi x + \zeta)$, with $\psi=J/Q^{3/2}$ the reduced current and $\zeta=G(\rho_0)$. This defines a master curve $\bar{\rho}(y)=G^{-1}(y)$ from which all density profiles follow after scaling space as $x=(y-\zeta)/\psi$, irrespective of the driving gradient or the average density. Moreover, temperature profiles scale now as $T(x)/Q=q[\rho(x)]^{-1}$, defining another master curve $\bar{T}(y)=q[\bar{\rho}(y)]^{-1}$. Note that similar scaling laws hold for any $d$-dimensional fluid with homogeneous interactions (including hard hyperspheres), see Appendix A.

As the density dependence of both the hard-disks EoS and conductivity are currently unknown, so are the scaling functions $\bar{\rho}(y)$ and $\bar{T}(y)$. However, we can measure them using the previous scaling scheme. To do so, we performed a large set of event-driven simulations of $N\in [1456,8838]$ hard disks of radius $\ell$ in a two-dimensional box of unit size $L=1$, with stochastic thermal walls \cite{fourier1} at $x=0,L$ at temperatures $T_0\in[2,20]$ and $T_L=1$, respectively, and periodic boundary conditions along the $y$-direction. The disks radius is defined by $N$ and the global packing fraction $\eta=\pi\ell^2N/L^2\in[0.05,0.8]$ via $\ell=\sqrt{\eta/N\pi}$, so that we can approach the $N\to \infty$, thermodynamic limit at constant, nonzero temperature gradient $\Delta T=|T_L-T_0|/L$ and fixed packing fraction. 

\begin{figure}[t]
\centerline{\includegraphics[width=8.5cm,clip]{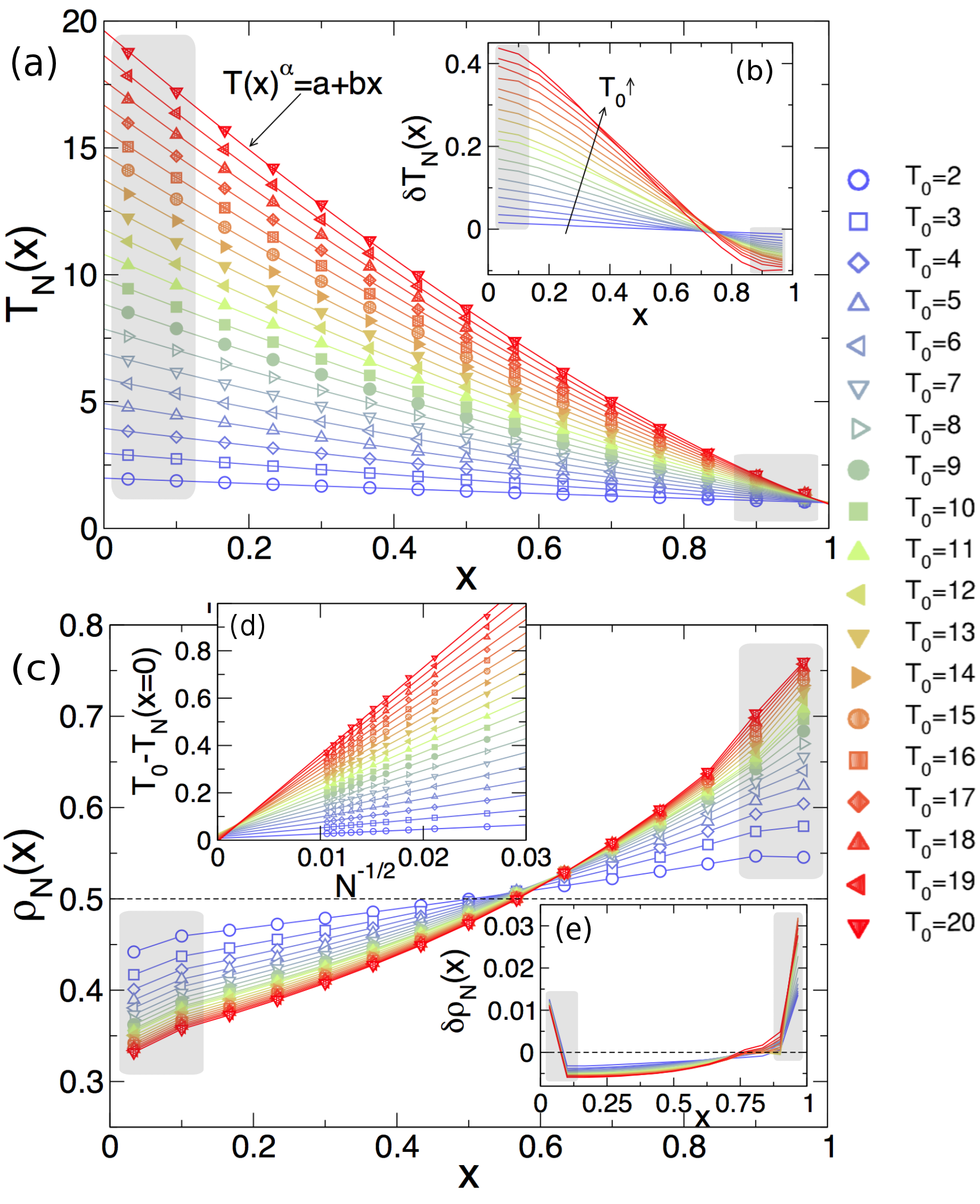}}
\vspace{-0.3cm}
\caption{\small (Color online)
(a) Temperature profiles for $N=8838$, $\eta=0.5$ and varying $T_0\in[2,20]$. Lines are nonlinear fits of the form $T(x)^\alpha=ax+b$ \cite{Eckmann}. Shaded (light grey) areas correspond to boundary
layers. (b) Finite size effects as captured by $\delta T_N(x)\equiv T_{N_{\text{max}}}(x) - T_{N_{\text{min}}}(x)$, with $N_{\text{max}}=8838$ and $N_{\text{min}}=1456$, for different gradients. (c) Density profiles for the same conditions that the top panel. (d) Thermal boundary resistance as a function of $N^{-1/2}$ for different $T_0$, and linear fits. (e) Finite size effects in density profiles, as captured by $\delta\rho_N(x)$, localize near the thermal walls. 
}
\label{fig2}
\end{figure}

We measured locally a number of relevant observables, including the local average kinetic energy, virial pressure, packing fraction, etc., as well as the heat current flowing through the thermal baths and the pressure exerted on the walls. Our time unit was set to one collision per particle on average, and time averages were performed with measurements every $10$ time units for a total time of $10^6-10^7$, after a relaxation time of $10^3$ time units which was empirically found sufficient to guarantee convergence to the steady state. For local measurements we divided the system into 15 virtual cells along the gradient direction, a fixed number of cells independent of the system parameters. Such discretization of the underlying continuous density and temperature profiles introduces some small corrections ($\sim 0.1\%$) that we explicitly take into account and  subtract (see Appendix B). Statistical errors in data averages were computed at a $99.7\%$ confidence level, and in most figures data errorbars are smaller than the plotted symbols (if not, errorbars are shown).

\begin{figure}[t]
\vspace{-0.5cm}
\centerline{\includegraphics[width=8.5cm,clip]{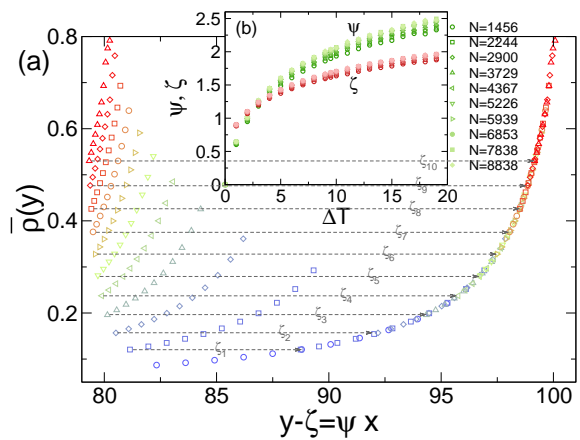}}
\vspace{-0.3cm}
\caption{\small (Color online) 
(a) Bulk density profiles for $N=2900$, $T_0=20$ and varying $\eta\in[0.15,0.65]$, as a function of $\psi x=Jx/Q^{3/2}$. By shifting each curve an amount $\zeta$, a perfect collapse is obtained which reconstructs the master
curve $\bar{\rho}(y)$. (b) Measured reduced currents $\psi$ and shifts $\zeta$ as a function of $\Delta T$ for different $N$ and $\eta=0.5$. Finite size effects are apparent.
}
\label{fig3}
\end{figure}

\begin{figure}[t]
\centerline{\includegraphics[width=8.5cm,clip]{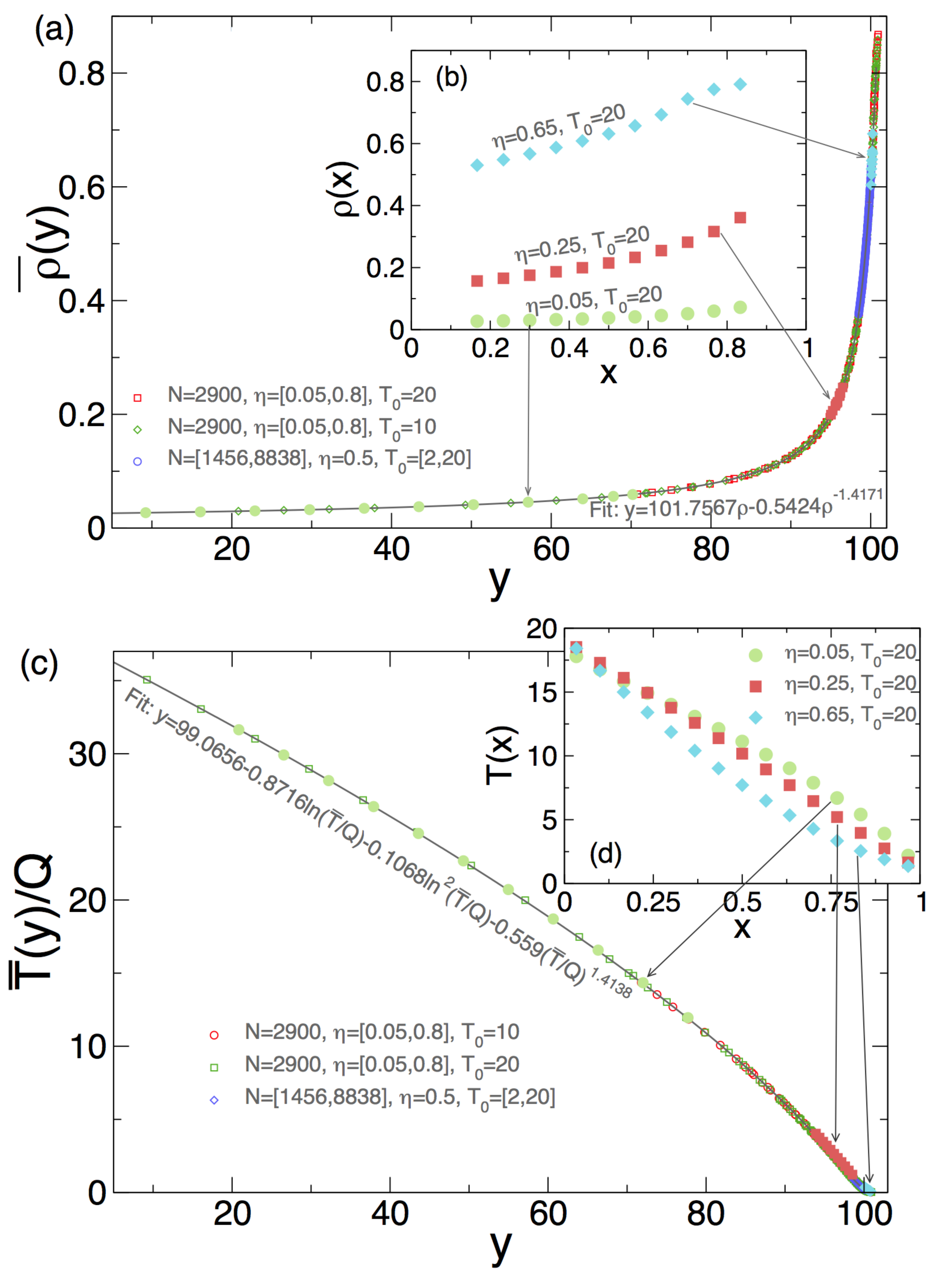}}
\vspace{-0.3cm}
\caption{\small (Color online) 
(a) Collapse of scaled bulk density profiles measured for $N\in[1456,8838]$ and different sets of conditions (see legend) for a total of more than 4000 data points. (b) Widely different bulk density
profiles measured for different conditions collapse onto different parts of the same master curve. (c)-(d) Collapse of bulk temperature profiles for the same conditions that the top panel. Note that the
shifts $\zeta$ obtained from the density scaling yield a perfect scaling for temperature profiles.
}
\label{fig4}
\end{figure}

\begin{figure}
\vspace{-0.55cm}
\centerline{\includegraphics[width=8.5cm,clip]{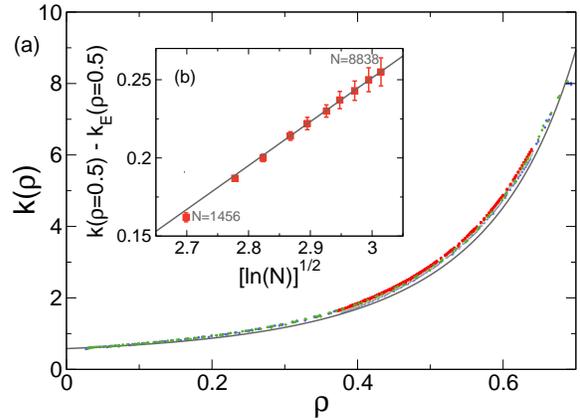}}
\vspace{-0.3cm}
\caption{\small (Color online)
(a) Density dependence of the heat conductivity as obtained from the rescaled temperature profiles $\bar{T}(y)\equiv T(y)/Q$ for different $\eta\in[0.05,0.8]$, $T_0\in[2,20]$ and $N\in[1456,8838]$. A well-defined deviation from Gass result $k_\text{E}(\rho)$ based on Enskog kinetic theory (full line) is found \cite{enskog,Gass}. Moreover, a systematic dependence with system size is also observed, see inset (b) for $\bar{\rho}=0.5$, which scales as $\sqrt{\ln(N)}$ for large enough $N$ \cite{tails}.
}
\label{fig5}
\end{figure}

Figs. \ref{fig2}.a,c show the temperature and density profiles measured for $N=8838$, $\eta=0.5$ and different gradients $\Delta T$, which are in general nonlinear. In all cases, the thermal
walls disrupt the structure of the surrounding fluid and this perturbation, most evident in density profiles, spreads toward the bulk of the system for a finite penetration depth, defining two
\emph{boundary layers} near the walls where finite size effects concentrate and become maximal, see Figs. \ref{fig2}.b and \ref{fig2}.e. The boundary disturbance also appears as a thermal
resistance or temperature gap between the extrapolated $T_N(x=0,L)$ and the bath temperature $T_{0,L}$ which decays as $N^{-1/2}$ for each $\Delta T$, see inset (d) in Fig. \ref{fig2}. In order
to perform the scaling analysis, we hence proceed to eliminate the boundary layers by removing from the profiles the two cells immediately adjacent to each wall
(see shaded areas in Fig. \ref{fig2}). The \emph{bulk profiles} $\rho(x)$ so obtained are then scaled using the reduced current $\psi=J/Q^{3/2}$ in each case (calculated by measuring the finite-size
heat current $J$ and reduced pressure $Q$) and shifted by a constant $\zeta$ to achieve a maximum overlap among all scaled profiles. Fig. \ref{fig3}.a shows an example of this scaling procedure for density profiles.

Using this method, we were able to collapse onto a single master curve $\bar{\rho}(y)$ a large amount of data for density profiles gathered for different $N$, $\Delta T$ and $\eta$, see Fig. \ref{fig4}.a.
Using the shifts $\zeta$ measured for density, all rescaled temperature profiles also collapsed onto another master curve $\bar{T}(y)$, see Fig. \ref{fig4}.c.
Strikingly, while the measured $J$, $Q$, $\psi$ and $\zeta$ depend on $N$ in a nontrivial way for each $\Delta T$ and $\eta$ (see Fig. \ref{fig3}.b), the collapsed data show no appreciable finite-size
effects, defining two master curves as predicted by the macroscopic theory. Such remarkable collapse thus implies that the measured bulk profiles are those of a \emph{macroscopic} hard-disk fluid 
obeying Fourier's law and subject to some renormalized, effective boundary conditions set by the boundary layers, which sum up all sorts of finite-size effects and boundary corrections. This
striking \emph{bulk-boundary decoupling phenomenon}, and the fine structural self-tuning of the fluid it involves (which goes beyond the mere presence of boundary layers), is even more surprising
at the light of the long range correlations present in nonequilibrium fluids \cite{maes,longrange}, and is likely to appear in most complex systems driven out of equilibrium by different boundary reservoirs,
offering a tantalizing method to avoid unreliable finite-size scaling extrapolations. In fact, a standard finite-size scaling analysis of our data, aimed at obtaining first the asymptotic ($N\to\infty$)
observables $\rho_{\infty}(x)$, $J_{\infty}$ and $Q_{\infty}$ for each $\Delta T$ and $\eta$ to perform then the scaling collapse, fails badly as none of these observables follow a clear asymptotic behavior.
In addition, the excellent scaling behavior of our data strongly suggests that, quite remarkably, Fourier's law (\ref{fourier}) remains empirically valid even under strong temperature gradients, extending 
its range of validity deep into the highly nonlinear regime. This means in particular that the higher order (Burnett) corrections conjectured for strong driving are in fact absorbed into the nonlinear 
conductivity $\kappa(\rho,T)$ in eq. (\ref{fourier}) \cite{fnote3}. The combination of our scaling analysis and the bulk-boundary decoupling phenomenon here described hence allows to obtain clean 
properties of macroscopic nonequilibrium fluids from finite-size simulations or experiments. The two master curves in Fig. \ref{fig4} have full predictive power, as we can deduce from them and the
scaling formulae in eqs. (\ref{JQcte})-(\ref{density}) the density and temperature profiles of a macroscopic hard disk system for any set of parameters $T_0$, $T_L$ and $\eta$. 

Our detailed data for the master curves in Fig. \ref{fig4} allow also for a precise measurement of the hard-disks heat conductivity over a broad range of densities. In fact, by multiplying
Fourier's law (\ref{fourier}) by $Q^{-3/2}$ and recalling the separable form of the conductivity, $\kappa(\rho,T)=\sqrt{T}k(\rho)$, it is easy to show that
$k(\rho)=[\sqrt{\bar{T}(y)} |\bar{T}'(y)|]^{-1}=J[\sqrt{T(x)} |T'(x)|]^{-1}$, with $\rho=\bar{\rho}(y)$. We hence performed discrete derivatives of the measured master curve $\bar{T}(y)$ for 
each of the different sets of parameters $\Delta T$, $\eta$ and $N$, identifying each value of $[\sqrt{\bar{T}(y)} \tilde{T}'(y)]^{-1}$ with the associated $\rho=\bar{\rho}(y)$. Fig. \ref{fig5}.a shows
the resulting $k(\rho)$, which exhibits deviations from the Gass prediction based on Enskog kinetic theory \cite{enskog,Gass}, as already reported \cite{brey,Risso}. Furthermore, a very 
weak but systematic $\sqrt{\ln N}$-dependence of $k(\rho)$ is observed, see inset (b) in Fig. \ref{fig5}, confirming for the first time and with high accuracy the marginally, $\sim\sqrt{\ln N}$ 
anomalous heat conductivity predicted for hard disks as a result of the long time tails in two dimensions \cite{tails}. This shows that our scaling method, together with the bulk-boundary
decoupling mechanism, allows one to get rid of artificial finite-size effects related with the presence of boundaries, which result in systematic errors in heat conductivity measurements,
keeping physically relevant bulk finite-size information.

In summary, we have shown that the nonequilibrium structure of a broad class of $d$-dimensional fluids obeys strikingly simple scaling laws when subject to a temperature gradient. We expect
similar, albeit more complex, scaling laws to hold in sheared fluids \cite{jpozo1}. We have measured the associated master curves in extensive simulations of hard disks, uncovering along the
way a remarkable bulk-boundary decoupling phenomenon by which all sorts of finite size effects and boundary corrections are renormalized into new boundary conditions on the remaining bulk fluid,
which obeys the macroscopic laws. The chances are that this subtle structural mechanism will also characterize the behavior of many real fluids with finite boundary layers. Finally, our scaling 
results remain valid under strong temperature gradients, extending the range of validity of Fourier's law deep into the highly nonlinear regime.

Financial support from Spanish projects FIS2009-08451 (MICINN) and FIS2013-43201-P (MINECO), University of Granada, Junta de Andaluc\'{\i}a projects P06-FQM1505, P09-FQM4682 and GENIL PYR-2012-1 and PYR-2014-13 projects is acknowledged.

\appendix

\section{Scaling for inverse power-law potentials in $d$-dimensions}
\label{appendixA}

The density-temperature separability of both the EoS and the heat conductivity is a main trait of hard disks which has proved particularly useful to understand their nonequilibrium scaling behavior starting from the local equilibrium and Fourier's law hypotheses. In particular, this property implies that the general master surfaces $\bar{\rho}_Q(y)$ and $\bar{T}_Q(y)$, from which any density and temperature profile follow arbitrarily far from equilibrium, in fact collapse onto a pair of universal master curves, $\bar{\rho}(y)$ and $\bar{T}(y)$. Here we show for completeness that such density-temperature separability is generic for $d$-dimensional fluids with pairwise inverse power-law (IPL) interactions, or IPL fluids in short, a property well-known in literature  (see e.g. Ref. [26] in the paper). Therefore we expect simplified scaling properties, similar to those of hard disks, to hold for this broad class of systems of both technological and fundamental importance. Such scaling laws may have direct applications for the physics of model glasses and other amorphous materials.

Inverse power-law potentials in $d$ dimensions take the following form
\begin{equation}
V(\vrr)=\epsilon\left(\frac{\sigma}{r}\right)^n
\label{ipl}
\end{equation}
where $r$ is the $d$-dimensional euclidean distance between two particles, while $\epsilon$ and $\sigma$ set the energy and length scales, respectively. Hard $d$-dimensional spheres are a particular case of IPL fluids in the $n\to\infty$ limit 
\be
V(r)=\left\{ \begin{array}{cc} 
0& \text{if  } r>\sigma \nonumber\\
\infty & \text{if  } r<\sigma
\end{array} \right. \, ,
\ee
where now $\sigma=2\ell$ with $\ell$ the radius of the hypersphere. We will show below that both the EoS and the heat conductivity of IPL fluids exhibit density-temperature separability. In particular, the IPL EoS can be written as
\be
P=\tilde\beta^{-1}q(\tilde\rho) \, ,
\label{eosipl}
\ee
with $P$ the pressure, while the IPL conductivity obeys
\be 
\kappa=\frac{\sigma^a\epsilon^b}{2m}\tilde\beta^c k(\tilde\rho) \, ,
\label{conducipl}
\ee
where we have defined the scaled inverse temperature $\tilde{\beta}$ and the scaled packing fraction $\tilde{\rho}$ as
\begin{eqnarray}
\tilde\beta&=&\beta \ell_{\text{eff}}^d \, , \nonumber\\
\tilde\rho&=&\rho\ell_{\text{eff}}^d \, ,
\end{eqnarray}
with $\ell_{\text{eff}}=\sigma(\beta\epsilon)^{1/n}$ an effective size for the soft particles. The nontrivial exponents in eq. (\ref{conducipl}) are
\begin{equation}
a=\frac{n(2-d)}{2(n+d)}\quad,\quad b=\frac{2-d}{2(n+d)}\quad,\quad c=\frac{2-2d-n}{2(n+d)} \, .
\end{equation}
The functions $q(\rho)$ and $k(\rho)$ are dimensionless, and $q(\rho)\simeq \rho$ in the ideal gas limit $\rho\simeq 0$. We now proceed to demonstrate the scaled density-temperature separability of eqs. (\ref{eosipl})-(\ref{conducipl}) for IPLs.

\subsection{Scaling form for the equation of state}
We first show that the canonical partition function of a system of $N$ particles in a volume $V$ at temperature $T$ interacting pairwise via the IPL potential (\ref{ipl}) obeys the following scaling relation
\begin{equation}
Z(N,V,T)=\left[\left(\frac{\beta}{2m}\right)^{1/2} \ell_{\text{eff}} \right]^{Nd}
\bar Z\left(N,\frac{V}{\ell_{\text{eff}}^d}\right) \, . \label{sca1}
\end{equation}
To prove this scaling, note that the canonical partition function $Z(N,V,T)$ is defined as
\begin{equation}
Z(N,V,T)=\frac{1}{N!h^{dN}}\int_V d\vec{r}^{(N)}\int_{\mathbb{R}^d}d\vec{p}^{(N)}\text{e}^{-\beta H(\vec{r}^{(N)},\vec{p}^{(N)})} \, ,
\label{Z}
\end{equation}
where $\vec{r}^{(N)}=(\vec{r}_1,\ldots,\vec{r}_N)$ and $\vec{p}^{(N)}=(\vec{p}_1,\ldots,\vec{p}_N)$ are the $2dN$ coordinates and momenta, respectively, $h$ stands for Planck's constant, and the Hamiltonian is given by
\begin{equation}
H(\vec{r}^{(N)},\vec{p}^{(N)})=\sum_{i=1}^N\frac{\vec{p}_i^2}{2m}+\epsilon\sigma^n\sum_{i<j}\frac{1}{\vert\vec{r}_i-\vec{r}_j\vert^n} \, .
\end{equation} 
We now change variables in the integrals of eq. (\ref{Z}) to scale the system parameters out of the exponential. In particular, by defining
\begin{equation}
\vec{u}_i=\sqrt{\frac{\beta}{2m}} \vec{p}_i\quad,\quad \vec{x}_i=\frac{\vec{r}_i}{\ell_{\text{eff}}} \, ,
\label{newvar}
\end{equation} 
we recover eq. (\ref{sca1}) with
\begin{equation}
\bar Z\left(N,\bar V\right)=\frac{1}{N!h^{dN}}\int_{\bar V} d\vec{x}^{(N)}\int_{\mathbb{R}^d}d\vec{u}^{(N)}\text{e}^{- \bar H(\vec{x}^{(N)},\vec{u}^{(N)})} \, ,
\end{equation}
where the parameter-free, scaled Hamiltonian reads
\begin{equation}
\bar H(\vec{x}^{(N)},\vec{u}^{(N)})=\sum_{i=1}^N\vec{u}_i^2+\sum_{i<j}\frac{1}{\vert\vec{x}_i-\vec{x}_j\vert^n}
\end{equation}

The equation of state can be now obtained from the canonical partition function as
\begin{equation}
P=\frac{1}{\beta}\frac{\partial}{\partial V}\ln Z(N,V,T)\biggr\vert_{N,T} \, .
\end{equation}
Using here the scaling form (\ref{sca1}) for $Z(N,V,T)$ we get
\begin{equation}
P=\beta^{-1}\ell_{\text{eff}}^{-d}\frac{\partial}{\partial\bar V}\ln\bar Z(N,\bar V)\biggr\vert_N \, ,
\end{equation}
where
$\bar V=V/\ell_{\text{eff}}^{d}$. The partial derivative of the rhs of the previous equation is necessarily a sole function of the density $\tilde \rho=N/\bar V = \rho \ell_{\text{eff}}^d$, so $\partial_{\bar V} \ln\bar Z(N,\bar V)\equiv q(\tilde \rho)$ and we recover the scaled density-temperature separable EoS of eq. (\ref{eosipl}) for IPL fluids.

\subsection{Scaling form for the thermal conductivity}
The thermal conductivity can be written via the Green-Kubo formula as the time integral of the energy current time correlation function measured in equilibrium, namely
\begin{equation}
\kappa=V\beta^2\int_0^\infty dt \langle J(0) J(t)\rangle_{eq} \, , 
\label{GK}
\end{equation} 
where we recall that units are chosen such that Boltzmann constant is set to one. The current is defined as
\begin{equation}
J=\frac{1}{mV}\sum_{i=1}^N\left[ \varepsilon_i p_{x,i}-\frac{1}{2}\sum_{j\ne i}(\vec{r}_{ij}\cdot\vec{p}_i)\frac{r_{x,ij}}{r_{ij}} V'(r_{ij})\right]
\end{equation}
where $r_{ij}=\vert \vec{r}_i-\vec{r}_j\vert$ and $\varepsilon_i=\vec{p}_i^2/2m+1/2\sum_{j\ne i} V(r_{ij})$ is the total energy of particle $i$. Moreover, we may write the current at time $t$ in terms of the current at time 0 as $J(t)=\exp(+t{\cal L})J(0)$, where we have used the system time evolution operator defined in terms of the system Liouvillian
\begin{equation}
{\cal L}b=\{b,H\}=\sum_{i,\alpha}\left[\frac{\partial H}{\partial p_{i\alpha} }\frac{\partial b}{\partial r_{i\alpha}}-\frac{\partial H}{\partial r_{i\alpha} }\frac{\partial b}{\partial p_{i\alpha}}\right] \, ,
\end{equation}
with $b$ an arbitrary dynamical function defined in phase space and $\{\cdot,\cdot\}$ the Poisson brackets. We may write now both the Liouvillian and the current in terms of the rescaled phase space variables $\vec{u}$ and $\vec{x}$ defined in eq. (\ref{newvar}). For the Liouvillian
\begin{equation}
{\cal L}=\frac{1}{\ell_{\text{eff}}\sqrt{2m\beta}}\bar {\cal L} \, ,
\end{equation}
with the definition
\begin{equation}
\bar {\cal L}=\sum_{i,\alpha}\left[ 2 u_{i\alpha}\frac{\partial}{\partial x_{i\alpha}}+n\sum_{j\ne i}\frac{x_{i\alpha}-x_{j\alpha}}{\vert \vec{x}_i-\vec{x}_j\vert^{n+2}}\frac{\partial}{\partial u_{i\alpha}}\right] \, .
\end{equation}
On the other hand, the current scales as
\begin{equation}
J=\frac{1}{V\beta \sqrt{2m\beta}}\bar J \, ,
\end{equation}
where we have defined
\be 
\bar J=\sum_{i=1}^N\left[2\bar \varepsilon_i u_{i,x}+n \sum_{j\ne i}(\vec{x}_{ij}\cdot\vec{u}_{ij})\frac{x_{ij,x}}{x_{ij}^{n+2}}\right] \, ,
\ee
with $\varepsilon_i=\beta^{-1} \bar \varepsilon_i$ and 
\begin{equation}
\bar \varepsilon_i=\vec{u}_i^2+\frac{1}{2}\sum_{j\ne i}\frac{1}{\vert\vec{x}_i-\vec{x}_j\vert^n}
\end{equation}
Substituting all these expressions in the Green-Kubo formula (\ref{GK}) for $\kappa$, we recover after some simple algebra the density-temperature separable scaling form of eq. (\ref{conducipl}) above for the thermal conductivity.

\subsection{Scaling for IPL systems}
The scaled density-temperature separability just demonstrated for IPL systems can be now used to write Fourier's law (\ref{fourier}) just in terms of the scaled density field in this more general case, similarly to what we did for hard disks,
\begin{equation}
\sqrt{2m}\left(1+\frac{d}{n}\right)\sigma^{\bar a}\epsilon^{\bar b} \, J P^{\bar c}=\bar G'(\tilde\rho)\frac{d\tilde\rho}{dx} = \frac{d\bar G'(\tilde\rho)}{dx} \, ,
\label{general}
\end{equation}
where $\bar G'(\tilde\rho)= k(\tilde\rho)q(\tilde\rho)^{\bar c-1}q'(\tilde\rho)$, and
\be
\bar a=-\frac{n(d+2)}{2(n+d)} \, , \, \bar b=-\frac{d+2}{2(n+d)} \, , \, \bar c=\frac{2-2d-3n}{2(n+d)} \, . \nonumber
\ee
This immediately implies the existence of a pair of master curves for IPL systems from which any steady state density and scaled temperature profiles follow, in the spirit of the hard disks result. Moreover, note that the hard disks results, or more generally the results for $d$-dimensional hard spheres, are recovered in the $n\to\infty$ limit.

\section{Discretization effects in density and temperature profiles}
\label{appendixC}

Once the hard disks system is driven to the stationary state, we measure the local temperature (i.e. local average kinetic energy) and local packing fraction at each of the $15$ cells in which we divide the simulation box along the gradient (i.e. $x$-) direction. When a disk overlaps with any of the imaginary lines separating two neighboring cells, it contributes to the density and kinetic energy of each cell proportionally to its overlapping area. The number of cells is fixed in all simulations to 15, independently of $N$, $\eta$, $T_0$ or $T_L$, so each cell becomes \emph{macroscopic} in the asymptotic thermodynamic limit. The local average of density and temperature around a finite neighborhood of a given point in space must be related with the underlying continuous profiles in order to subtract any possible bias or systematic correction from the data.

Let's $T_C$ and $\rho_C$ be the temperature and packing fraction in a cell centered at $x_c\in[0,L]$ of size $\Delta$. Assuming that there exist continuous (hydrodynamic) temperature and density profiles $T(x)$ and $\rho(x)$, we can relate the cell averages to the continuos profiles by noting that
\ben
T_C&=&\frac{1}{\Delta\rho_C}\int_{x_c-\Delta/2}^{x_c+\Delta/2}dx\,\rho(x)T(x) \, , \nonumber \\
\rho_C&=&\frac{1}{\Delta}\int_{x_c-\Delta/2}^{x_c+\Delta/2}dx\,\rho(x) \, . \nonumber 
\een
We may expand now the continuous profiles around $x_c$ inside the cell of interest and solve the above integrals. Keeping results up to $\Delta^2$ order
\begin{eqnarray}
T_C&=&\frac{1}{\rho_C}\biggl[\rho(x_c)T(x_c)+\frac{\Delta^2}{24}\frac{d^2}{dx^2}\left[\rho(x)T(x)\right]_{x=x_c}+O(\Delta^3)\biggr]\, , \nonumber\\
\rho_C&=&\rho(x_c)+\frac{\Delta^2}{24}\frac{d^2\rho(x)}{dx^2}\vert_{x=x_c}+O(\Delta^3) \, .\nonumber
\end{eqnarray} 
By inverting the above expressions, we arrive to the desired result, namely
\begin{eqnarray}
T(x_c)&=&T_C-\frac{1}{24}\biggl[\frac{2}{\rho_C}(\rho_{C+1}-\rho_C)(T_{C+1}-T_C) \nonumber \\
&+&T_{C+1}-2T_C+T_{C-1} \biggr]\label{TC}\\
\rho(x_c)&=&\rho_C-\frac{1}{24}\left[\rho_{C+1}-2\rho_C+\rho_{C-1}\right]\label{rhoC}
\end{eqnarray}
Typically these corrections to the cell density and temperature are small ($\sim 0.1\%$), but they turn out to be important for disentangling the different finite size effects in order to obtain the striking collapse of measured density and temperature profiles onto the master curves described in the main text.


\begin{thebibliography}{99}
\bibitem{conds} The boundary conditions for the density field can be inferred from the constraints
\be
\phantom{aaa} \frac{T_0}{T_L} = \frac{f_Q(\rho_0)}{f_Q(\rho_L)}  \quad ; \quad
\eta =  \frac{1}{L}\int_0^L \rho(x) dx = \frac{\displaystyle \int_{\rho_0}^{\rho_L} \rho \, G'_Q(\rho) d\rho}{\displaystyle G_Q(\rho_L)-G_Q(\rho_0)} \, . \nonumber
\ee
These equations, together with eq. (\ref{JQcte}) above, completely define the macroscopic problem in terms of $\rho(\vrr)$. The reduced pressure and the heat current can be now obtained as $Q=q(\rho_0,T_0)$ and $J=[G_Q(\rho_L)-G_Q(\rho_0)]/L$.
\bibitem{exist} We assume that $G_Q(\rho)$ has a well-defined inverse $G_Q^{-1}(y)$. This assumption seems reasonable as steady density profiles are typically well behaved.
\end{thebibliography}
\end{document}